# Deep-Learning-Enabled Inverse Engineering of Multi-Wavelength Invisibility-to-Superscattering Switching with Phase-Change Materials

**JIE LUO[1], XUN LI[1], XINYUAN ZHANG[2,3], JIAJIE GUO[2,3], WEI LIU[4], YUN LAI[5], YAOHUI ZHAN[2,3,6], MIN HUANG[2,7]**

[1]*School of Physical Science and Technology, Soochow University, Suzhou 215006, China*
[2]*School of Optoelectronic Science and Engineering & Collaborative Innovation Center of Suzhou Nano Science and Technology, Soochow University, Suzhou 215006, China.*
[3]*Key Lab of Advanced Optical Manufacturing Technologies of Jiangsu Province & Key Lab of Modern Optical Technologies of Education Ministry of China, Soochow University, Suzhou 215006, China.*
[4]*College for Advanced Interdisciplinary Studies, National University of Defense Technology, Changsha, Hunan 410073, China*
[5]*National Laboratory of Solid State Microstructures, School of Physics, and Collaborative Innovation Center of Advanced Microstructures, Nanjing University, Nanjing 210093, China*
[6]*e-mail:* [yhzhan@suda.edu.cn](mailto:yhzhan@suda.edu.cn)
[7]*e-mail:* [hmin@suda.edu.cn](mailto:hmin@suda.edu.cn)

**Abstract:** Inverse design of nanoparticles for desired scattering spectra and dynamic switching between the two opposite scattering anomalies, i.e. superscattering and invisibility, is important in realizing cloaking, sensing and functional devices. However, traditionally the design process is quite complicated, which involves complex structures with many choices of synthetic constituents and dispersions. Here, we demonstrate that a well-trained deep-learning neural network can handle these issues efficiently, which can not only forwardly predict scattering spectra of multilayer nanoparticles with high precision, but also inversely design the required structural and material parameters efficiently. Moreover, we show that the neural network is capable of finding out multi-wavelength invisibility-to-superscattering switching points at the desired wavelengths in multilayer nanoparticles composed of metals and phase-change materials. Our work provides a useful solution of deep learning for inverse design of nanoparticles with dynamic scattering spectra by using phase-change materials.

## 1. Introduction

The study of light scattering by nanoparticles has a long history [1-5] and is of great importance to many fields of physics and materials science. Recently, increased interest in optical imaging, sensing, and energy harvesting at nanoscales has attracted renewed interest in the research of light scattering on the capability of engineering scattering spectra at will. Control of light scattering is shown in two extreme scattering anomalies: superscattering [6-16] and invisibility (cloak) [17-28]. Superscattering, which demonstrates an extraordinarily large scattering cross section (CS) far greater than that of the single-channel limit, has practical significance for applications such as imaging, biomedicine, and photovoltaics [3]. In theory, an arbitrary large superscattering can be enabled by the orthogonality of channels if a sufficient number of channels (e.g. cylindrical and spherical harmonics that characterize multipoles of different orders) are excited for degenerate resonances [29]. On the other hand, the total scattering CS of a particle can be dramatically reduced with plasmonic or dielectric multilayer structures [23, 24]. In this regime, excited multipoles inside particles destructively interfere with each other in the far-field, so that the particle does not scatter light to any channel [30].

Dynamic control of light scattering between the two opposite scattering anomalies (i.e. superscattering and invisibility) at nanoscale dimensions is highly desired in advanced and intelligent optical devices. Recently, phase-change materials (PCMs) [9, 31-41], which can switch between different crystalline states as a function of an external bias, are becoming one of the most promising approaches to dynamically and efficiently control light at subwavelength scales. The phase transition can be induced by applying either heat [32-35], photon [36, 37], or electric energy [42] on the PCMs, offering a pronounced change of their dielectric function. With this unique property of PCMs, switch between superscattering and invisibility has been proposed recently in core-shell nanoantennas composed of PCM germanium telluride (GeTe) and silver (Ag) [9]. However, due to the limited degree of freedom of design space, the interesting switching phenomenon is only observed at a particular wavelength. Although a multilayer configuration in principle can provide potential possibility toward multi-wavelength invisibility or superscattering [13, 14, 21], the multi-wavelength switching has never been reported before as the design process would be quite complicated when taking dispersions and phase changing of constitutions into consideration.

Nowadays, design and optimization approaches based on deep learning are rapidly emerging, where artificial neural networks [43, 44] are trained to overcome complicated design problems in advanced nanophotonic and electromagnetic devices with growth of structural complexity and higher degree of freedom. A trained neural network can be used as a fast, general purpose predictor of optical and electromagnetic responses of complicated three-dimensional structures [45]. Through embedding with trained neural networks to rapidly interpret measured data and co-design the measurement setup with reconfigurable elements, intelligent metadevices can be designed, such as reconfigurable metasurface transceivers [46], metasurface-based electromagnetic sensors [47], self-adaptive cloak [48], etc. Notably, neural networks are particularly efficient in solving notoriously difficult inverse problems in nanophotonics in comparison with traditional approaches like genetic algorithm [49]. Recently, neural networks have shown great feasibility in the inverse design of various nanostructures with desired optical responses, including nanoparticles [49, 50], multilayer films [51, 52], metasurfaces [53-58], waveguide systems [59], etc. In addition, deep learning based on photonic platforms instead of conventional computers with von Neumann architecture can further improve computational speed and power efficiency [60, 61]. These works provide us powerful tools to realize the interesting behavior of multi-wavelength superscattering-to-invisibility switching.

In this work, we propose a deep-learning-assisted method to inversely design hybrid PCM-metal multilayer nanoparticles for multi-wavelength invisibility-to-superscattering switching at the desired wavelengths. Here, we show that an artificial neural network can predict the scattering spectra of hybrid PCM-metal multilayer nanoparticles with high precision after the training on a small sampling of data. Once trained, our neural network can solve inverse design problems and predict the required structural parameters (including radius and number of layers) and material information (including the selection of metal and the phase of PCM) efficiently for given scattering spectra. Furthermore, we demonstrate the inverse design of a hybrid GeTe-Ag nanoparticle for triple-wavelength invisibility-to-superscattering switching. We show that superscattering is simultaneously obtained at 1785nm, 2180nm and 2435nm under low temperatures as the result of spectral overlap of dipole and quadrupole resonances, while the scattering CS of all the three wavelengths will become negligibly small when the temperature is high. Our work would provide a useful solution of deep learning for inverse design of nanoparticles with dynamic scattering spectra by using PCMs.

## 2. DEEP-LEARNING-ENABLED FORWARD PREDICTION AND INVERSE DESIGN

We consider a multilayer nanoparticle composed of a metal core (radius $t_1$) and 5 alternating shells of PCM GeTe and metal (thickness $t_2$, $t_3$, $t_4$, $t_5$ and $t_6$), as illustrated by Fig. 1(a). Here, we have three choices of possible metals, that is, Ag, gold (Au) and aluminum (Al), whose dispersions are taken from Ref. [62]. The GeTe belongs to the class of PCMs possessing the ability of rapidly switching between different crystalline phases. The phase transition from amorphous phase to crystalline would happen when temperature is increased to ~420K, thus leading to a substantial change in its dielectric constant. The refractive index ($n$) and extinction coefficient ($\kappa$) of the GeTe under low and high temperatures [9, 31] are plotted in Figs. 1(c) and 1(d), respectively. By using laser irradiation, the change in optical properties on the transition between the amorphous and crystalline states can be realized on the nanosecond timescale (<100ns) [63].

The proposed multilayer nanoparticle can be represented by an array $\{t_1, t_2, t_3, t_4, t_5, t_6; m\}$, where $m$ represents the material type (including the choices of metals and the phase of GeTe). We note that the real structure can have shells less than 5 if there are zero values in the array $\{t_2, t_3, t_4, t_5, t_6\}$.

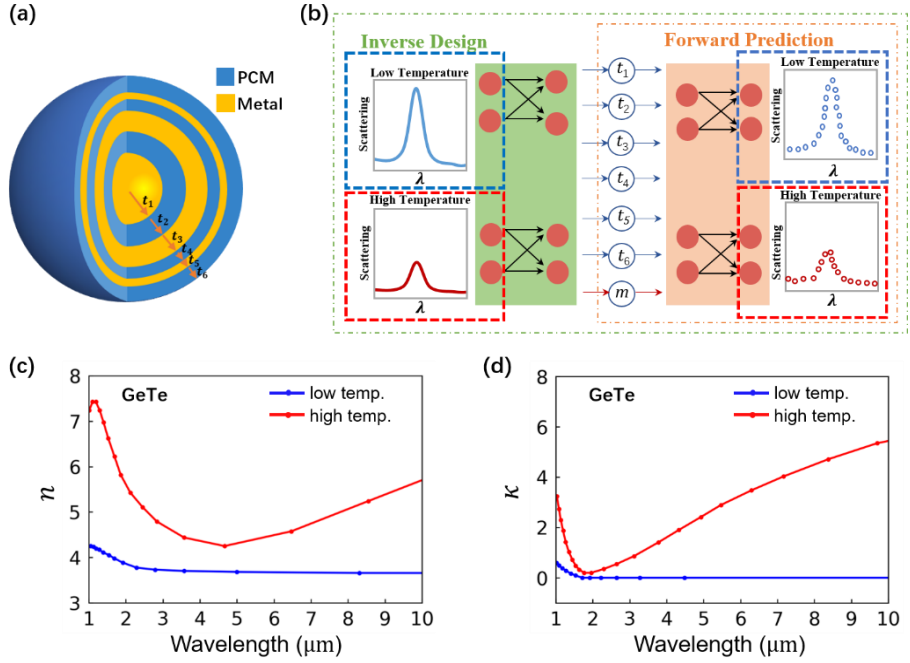

**Fig. 1.** (a) Schematic of a hybrid PCM-metal multilayer nanoparticle. (b) Illustration of the artificial neural network architecture for forward prediction and inverse design. (c) Refractive index and (d) extinction coefficient of the GeTe under low (blue lines) and high (red lines) temperatures.

For any given array of parameters, the scattering spectra of this multilayer nanoparticle can be calculated analytically based on Mie scattering theory, which provides rigorous solutions for the optical scattering by spherical particles [1, 2]. According to the Mie theory, the scattering CS of a particle under plane wave incidence can be expressed as

$$Q = \frac{\lambda^2}{2\pi} \sum_{l=1} (2l+1)\left(\left|c_l^E\right|^2 + \left|c_l^M\right|^2\right), \qquad (1)$$

where $\lambda$ is the wavelength in free space. The quantity $l$ defines the order of the scattering channel, which is related to angular momenta. $c_l^E$ and $c_l^M$ denote the scattering coefficients for transverse-electric (magnetic multipolar) and transverse-magnetic (electric multipolar) modes, respectively.

Generally, the scattering CS of a subwavelength nanoparticle is dominated by the dipole term ($l=1$), so that only $c_1^E$ and $c_1^M$ govern the scattering properties [2]. When $\left|c_1^E\right|=1$ or $\left|c_1^M\right|=1$, the scattering resonance occurs. In this case, the nanoparticle scatters the maximum possible scattering associated with the given harmonic. In passive systems, the maximum scattering CS contribution from any given harmonic is $(2l+1)\lambda^2/2\pi$, which leads to $3\lambda^2/2\pi$ for a subwavelength nanoparticle with dipole resonance ($l=1$). This limit is referred to as the single-channel limit [6]. Superscattering of a nanoparticle will occur when the total scattering CS is larger than $3\lambda^2/2\pi$ [6-9].

By using the Mie theory, the original data sets are generated with 72000 entries. Each data sample is a typical structural data with the vector of the quantized material as well as the corresponding spectral information. Eight parameters (6 layer's radius/thickness value, 1 code of material type and 1 code of the phase/temperature) are adopted for the material information, while 401 discrete points within 1000nm to 3000nm are directly utilized to construct the vector of its spectra. Original data samples are firstly shuffled and then the training dataset with 57600 entries (80%) are randomly picked up from the samples. The left data samples with 14400 entries (20%) form the evaluation dataset. The back propagation process of the neural network (Fig. 1(b)) will try to minimize the discrepancy between predicted results by the network and the realistic values, and then update parameters of the network (e.g. weights, biases, etc.) until the loss reaches to the acceptable status.

The structure of the forward neural network is 9-100-1200-1000-802, which is related to the number of neurons in 5 layers. The Leaky ReLu function is used as the activation function. The AdamW optimizer with a learning rate of 5e-4 and a weight decay of 1e-5 is exploited by setting the batch size to be 128. In practice, we have also tried other number of layers. When the number of layers is gradually increased, the loss decreases, while the runtime (including training time and forward computation time) increases dramatically. We find that the loss stops declining and reaches a stable value when the number of layers is increased to a certain extent (i.e. 5). Therefore, considering the balance of runtime and loss, we have chosen a 5-layered forward neural network.

We note that the forward model in principle can be directly based on the Mie scattering theory algorithm, which however is only valid for spherical nanoparticles. In order to ensure the generality of our algorithm in dealing with scattering spectra of nanostructures besides spherical nanoparticles, the neural network is used in the forward prediction.

The structure of the inverse neural network is 802-500-2000-2000-100-9, which corresponds to the number of neurons in 6 layers. The activation function, optimizer and batch size are same as those in the forward neural network. To realize the inverse design, we first train the forward neural network and fix its internal weights, which maps the design space to the response space. Then, we connect the inverse neural network to the left of the forward neural network to form a tandem network, which can map the response space to the response space. When the training is completed, predicted results can be obtained from the middle of the tandem network by feeding the scattering spectra to the input of the tandem network. All codes can be found at https://github.com/xiangxiangbest/PCMnet.git.

It is mentioned that, compared with [49] in which only structural parameters can be inverse designed, our neural network is capable of simultaneously designing structural parameters and material information. Unlike the algorithm in [50], the material information is labeled within the training data and predicted directly as a regression problem rather than a classification problem. This can avoid using the weighted loss function of material and structural parameters, thus simplifying the algorithm.

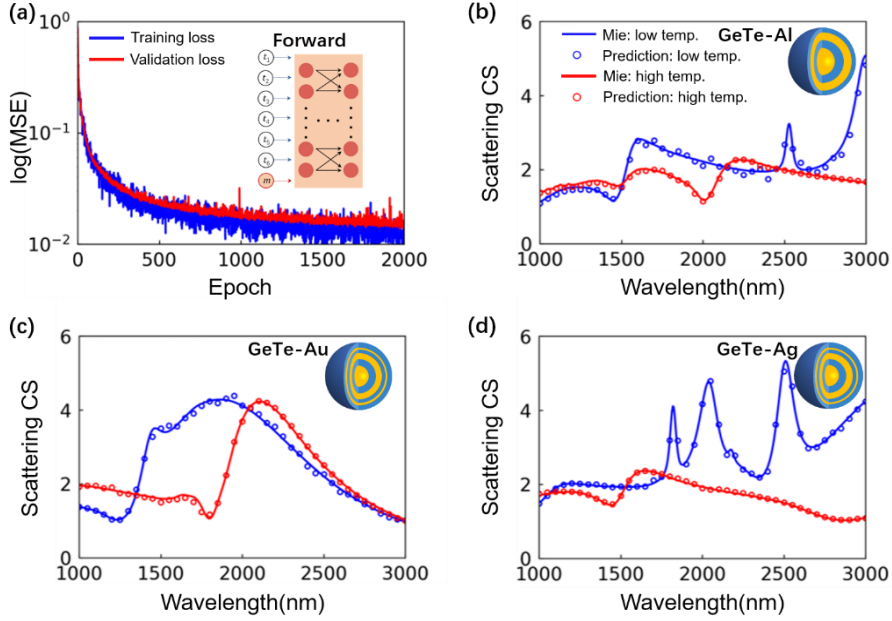

**Fig. 2.** (a) Training loss (blue) and validation loss (red) over epochs during the forward prediction. [(b)-(d)] Scattering CS of three test examples predicted by the deep neural network (dots) and Mie theory (lines) under high (red) and low (blue) temperatures. The model in (b) is a 4-layered GeTe-Al nanoparticle (i.e. $t_5 = t_6 = 0$), (c) is a 6-layered GeTe-Au nanoparticle, and (d) is a 6-layered GeTe-Ag nanoparticle.

Here, we choose the mean square error (MSE) as the loss function to evaluate the efficiency of the training and validation process. The MSE is identified as

$$L = \frac{1}{MN}\sum_{m=1}^{M}\sum_{n=1}^{N}(q_{n,m} - q'_{n,m})^2, \qquad (2)$$

where $q$ and $q'$ denote the quantized actual scattering CS and the network output result, respectively. $M$ and $N$ are the set epochs and the number of points of the quantized spectra, respectively. From Fig. 2(a), we find that both training loss (blue) and validation loss (red) decrease rapidly after only 500 epochs, demonstrating that the training is highly efficient. We shall note that if the number of epochs is too large, the neural network may overfit on the training dataset, accompanying with unstable or even increased validation loss. Therefore, during the training, the validation loss is evaluated after each epoch. When it stops declining over successive 10 epochs, we stop the training in order to avoid overfitting. Furthermore, regularization and dropout operations are exploited to further prevent overfitting.

For confirmation of the effectiveness of our neural network, three examples of randomly chosen test results are shown in Figs. 2(b)-2(d), which are related to a 4-layer GeTe-Al nanoparticle (i.e. $t_5 = t_6 = 0$), a 6-layer GeTe-Au nanoparticle, and a 6-layer GeTe-Ag nanoparticle, respectively. The red and blue lines/dots denote the cases under high and low temperatures, respectively. It is seen that the predicted scattering spectra by our network (dots) match Mie solutions (solid lines) well under both high and low temperatures. The losses in Figs. 2(b), 2(c) and 2(d) are found to be 0.0081, 0.0036 and 0.0036, respectively. These results demonstrate that our deep neural network is well trained with high precision. We note that a forward prediction costs only ~0.001 seconds in our workstation with dual 2080ti graphic card.

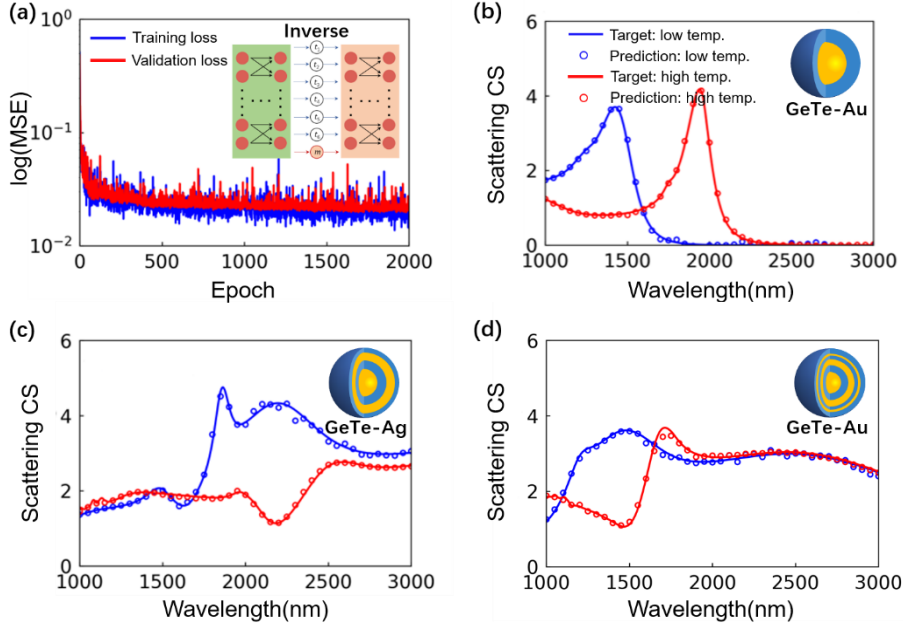

**Fig. 3.** (a) Training loss (blue) and validation loss (red) over epochs during the inverse design. [(b)-(d)] Comparison between three test examples of target spectra (lines) and predicted spectra by the deep neural network (dots) under high (red) and low (blue) temperatures. The insets show the predicted structures: (b) 2-layer GeTe-Au nanoparticle (i.e. $t_3 = t_4 = t_5 = t_6 = 0$ ), (c) 4-layer GeTe-Ag nanoparticle (i.e. $t_5 = t_6 = 0$ ), (d) 6-layer GeTe-Au nanoparticle.

|  | $t_1$(nm) | $t_2$(nm) | $t_3$(nm) | $t_4$(nm) | $t_5$(nm) | $t_6$(nm) | $m$ | Loss |
|---|---|---|---|---|---|---|---|---|
| Fig. 3(b) | 183.62 | 77.77 | 0 | 0 | 0 | 0 | Au | 0.027 |
| Fig. 3(c) | 77.2 | 234.8 | 263.05 | 80.93 | 0 | 0 | Ag | 0.040 |
| Fig. 3(d) | 114.49 | 92.1 | 29.18 | 57.7 | 124.88 | 33.79 | Au | 0.046 |

**Table 1.** Design parameters of predicted nanoparticles in Fig. 3.

Besides the fast and precise forward prediction, our trained network can also be used to solve nanoparticle inverse design problems. The process of inverse design through the network is illustrated in Fig. 1(b). Reverse network could make use of forward network, which has been fully trained before, to iterate repeatedly until the discrepancy between the predicted spectra and the realistic spectra reaches to an optimized status. In the following, we shall demonstrate the capability of our deep neural network to inversely design nanoparticles for given scattering spectra.

We randomly select target spectra from the sampling data that have never been used in previous training, so that the target spectra are physically realizable. The training loss (blue) and validation loss (red) over epochs are plotted in Fig. 3(a), showing that the inverse design network is well trained after 500 epochs. With this network, we test three target spectra, as plotted by solid lines in Figs. 3(b)-3(d). The red and blue lines denote the cases of high and low temperatures, respectively. Through the inverse design, the network finds out the correct structures and materials, that is, 2-layer GeTe-Au nanoparticle in Fig. 3(b), 4-layer GeTe-Ag nanoparticle in Fig. 3(c), and 4-layer GeTe-Au nanoparticle in Fig. 3(d), as illustrated by the insets. The detailed parameters can be found in Table 1. The dots denote the spectra predicted by the network, which show good agreement with target input spectra. These results clearly confirm the effectiveness of our neural network in inverse design. Here, it only takes ~0.03 seconds for each inverse design problem, much faster than traditional optimization techniques like genetic algorithms, which may take hours to solve the inverse design problem.

## 3. DEEP-LEARNING-ENABLED INVERSE ENGINEERING OF INVISIBILITY-TO SUPERSCATTERING SWITCHING

We know that the dynamic control of light scattering between the two opposite extreme phenomena (i.e. superscattering and invisibility) is of great interest to cloaking [19], sensing [64], and functional devices [65]. However, simultaneous spectral tuning under high and low temperatures is not easy. In particular, realizing multi-wavelength invisibility-to-superscattering switching in nanoparticles is quite difficult, if not impossible, as dispersions of different metals should be synthetically considered. Interestingly, here we show that our trained network can easily find out solutions through inverse design, as we shall demonstrate as follows.

The neural network we applied is illustrated in Fig. 4(a). The input is our optimization goal, that is, $\max|Q_L - Q_H|$ at chosen short wavelength ranges centered at $\lambda_1$, $\lambda_2$, $\lambda_3$... Here, $Q_L$ and $Q_H$ are the scattering CS at low and high temperatures, respectively. The output is the structural and material parameters of nanoparticles. In such an inverse design problem involving the optimization goal $\max|Q_L - Q_H|$, the trained forward neural network is used. We set a custom function at the output (i.e. $|Q_L - Q_H|$) with fixed weights, and set the input as a variable. The backpropagation algorithm to train the input of the network through continuous iteration is utilized. Then, the neural network can predict nanoparticles with appropriate structural parameters and material choices, which can satisfy the maximization of the custom function (i.e. $\max|Q_L - Q_H|$).

As an example, we select three wavelength ranges 1700-1900nm, 2000-2200nm and 2300-2500nm, which are centered at $\lambda_1 = 1800\,\text{nm}$, $\lambda_2 = 2100\,\text{nm}$ and $\lambda_3 = 2400\,\text{nm}$, respectively. The predicted nanoparticle is a 2-layer GeTe-Ag one. The detailed parameters can be found in Table 2. Here, to evaluate the output results, a loss function is defined as $\frac{\text{ave}|Q_L - Q_H|_{\text{outside}}}{\text{ave}|Q_L - Q_H|_{\text{inside}}}$, where $\text{ave}|Q_L - Q_H|_{\text{outside}}$ and $\text{ave}|Q_L - Q_H|_{\text{inside}}$, respectively, denote the averaged $|Q_L - Q_H|$ outside and inside the chosen wavelength ranges. A small loss indicates a large contrast of scattering CSs under low and high temperatures within the chosen wavelength ranges, while a low contrast outside the chosen wavelength ranges. Here, the loss is found to be 0.11, showing the effectiveness of our neural network.

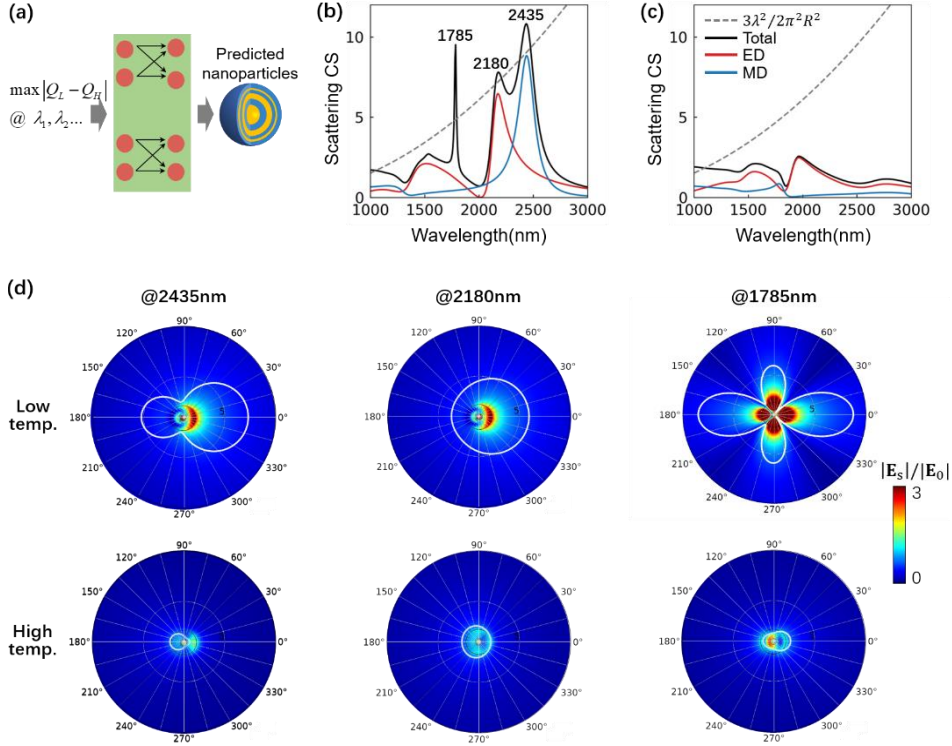

**Fig. 4.** (a) Illustration of the artificial neural network for inverse design of nanoparticles with multi-wavelength invisibility-to-superscattering switching. [(b) and (c)] Scattering spectra of the predicted nanoparticle under (b) low, (c) high temperature. The black, red and blue lines denote the total scattering CS, ED resonance and MD resonance, respectively. The gray dashed line is related to the single-channel limit. (d) Scattering far-field patterns (solid curves) and near-field distributions $|\mathbf{E}_s|/|\mathbf{E}_0|$ (color map) at wavelengths 2435nm (left), 2180nm (middle) and 1785nm (right) under low (upper) and high (lower) temperatures.

|  | $t_1$(nm) | $t_2$(nm) | $t_3$(nm) | $t_4$(nm) | $t_5$(nm) | $t_6$(nm) | $m$ | Loss |
|---|---|---|---|---|---|---|---|---|
| Fig. 4 | 74.69 | 241.51 | 0 | 0 | 0 | 0 | Ag | 0.11 |

**Table 2.** Design parameters of predicted nanoparticles in Fig. 4.

The total scattering CS of this nanoparticle is calculated based on Mie theory, as plotted by black solid lines in Fig. 4(b) (low temperature) and Fig. 4(c) (high temperature). It is seen from Fig. 4(b) that three scattering peaks occurring at 1785nm, 2180nm and 2435nm exceed the single-channel limit (gray dashed lines, $3\lambda^2/2\pi^2 R^2$ with $R$ being the radius of the whole particle), revealing the multi-wavelength superscattering phenomena under low temperature. Interestingly, from Fig. 4(c), we find that the scattering CS regarding to the three wavelengths is small. These results demonstrate the multi-wavelength switch between superscattering and near-invisibility.

In order to explore the underlying physics, in Figs. 4(b) and 4(c), we also plot the scattering CS for the electric-dipole (ED, red lines) and magnetic-dipole (MD, blue lines) modes, which are related to the scattering coefficients $c_1^E$ and $c_1^M$ in Eq. (1), respectively. We find that the superscattering behaviors at 2180nm and 2435nm are the result of the spectral overlap of ED and MD resonances. For verification, the scattering far-field patterns (solid curves) and near-field distributions $|\mathbf{E}_s|/|\mathbf{E}_0|$ (color map, $\mathbf{E}_s$ and $\mathbf{E}_0$ are the scattering and incident electric fields) are calculated by using software COMSOL Multiphysics, as displayed in Fig. 4(d). Evident strong dipole radiation is seen under low temperature (upper insets), while the scattering is quite small when the temperature is high (lower insets). Intriguingly, besides the contribution from the dipole resonances, we find that quadrupole resonances also contribute significantly. Indeed, the superscattering at 1785nm is induced by the quadrupole resonances, which is confirmed by the scattering far-field pattern and near-field distribution under low temperature (upper insets in Fig. 4(d)). Similarly, the scattering becomes negligibly small when the temperature increases. Therefore, we have shown the capability of our neural network in inverse design of nanoparticles with multi-wavelength invisibility-to-superscattering switching behaviors.

## 4. DISCUSSION AND CONCLUSION

In typical inverse design problems, one of the main challenges is non-uniqueness of the solutions [66, 67]. That is, meaningfully different design parameters could be predicted for a given scattering spectrum. Fortunately, the non-uniqueness issue is not severe in this work. Nanoparticles with meaningfully different structural parameters and materials would generally have very different scattering spectra. During the inverse design, the forward neural network is used to predict the spectra of the predicted nanoparticles by the inverse neural network, so that we can compare the output spectra with the target spectra and calculate the loss. Through minimizing the discrepancy, we finally will obtain a unique optimized nanoparticle, whose spectra match best with the target spectra.

Currently, our deep neural network is a good interpolator [55], which can only process data inside the training hyperspace well. However, we believe that through feeding the neural network with sufficient source data with rich features and transferring knowledge from other nanophotonic problems to the specific nanoparticle scattering problem [68], our neural network could also be capable of interpolating and extrapolating in the future.

In summary, based on the scattering switching problem in hybrid PCM-metal multilayer nanoparticles, we build an artificial neuron network. After the training on a small sampling of data, the network can realize not only the precise forward prediction but also the fast inverse design. In particular, we utilize this neuron network to overcome the design problem of multi-wavelength invisibility-to-superscattering switching. Through giving switching phenomena at chosen wavelengths, the network can predict the required materials (e.g. Au, Ag, Al), structural parameters (e.g. layer number and thickness) and phase states (e.g. high and low temperatures).

Finally, it is noteworthy that different from the experience-driven design method, here a data-driven method is employed for complicated scattering engineering. This method can not only significantly lower the technical barriers of physics knowledge, but also dramatically raise the possible complexity degree of design space. Through adding more material selections and introducing more complicated structures to enrich the training data, the developed neural network in principle can inversely design nanoparticles with almost arbitrarily given scattering spectra. Besides the scattering switching phenomena discussed above, many other attractive scattering spectra (e.g. Fano and Lorentzian profiles) can also be inversely designed through this neural network. We believe that our work opens interesting opportunities for inverse design of nanoparticles with attractive scattering spectra.

**Funding.**

National Natural Science Foundation of China (11704271, 61802272, 11874426, 61671314, 11974176), Natural Science Foundation of Jiangsu Province (BK20181167, BK20180834), Opening Project of State Key Laboratory of High Performance Ceramics and Superfine Microstructure (SKL201912SIC), Foundation of Equipment Development Department (6140922010901), Priority Academic Program Development of Jiangsu Higher Education Institutions.

**Acknowledgment.**

The authors acknowledge helpful discussions with Sergei Lepeshov on the scattering calculation.

**Disclosures.**

The authors declare no conflicts of interest.